# A Limited Habitable Zone for Complex Life

Edward W. Schwieterman[1,2,3,4,5], Christopher T. Reinhard[3,4,6], Stephanie L. Olson[3,7], Chester E. Harman[4,8,9], and Timothy W. Lyons[1,3,4]  
[1] Department of Earth and Planetary Sciences, University of California, Riverside, CA, USA; eschwiet@ucr.edu  
[2] NASA Postdoctoral Program, Universities Space Research Association, Columbia, MD, USA  
[3] NASA Astrobiology Institute, Alternative Earths Team, Riverside, CA, USA  
[4] Nexus for Exoplanet System Science (NExSS) Virtual Planetary Laboratory, Seattle, WA, USA  
[5] Blue Marble Space Institute of Science, Seattle, Washington, USA  
[6] School of Earth and Atmospheric Sciences, Georgia Institute of Technology, Atlanta, Georgia, USA  
[7] Department of Geophysical Sciences, University of Chicago, Chicago, IL, USA  
[8] NASA Goddard Institute for Space Studies, New York, NY, USA  
[9] Department of Applied Mathematics and Applied Physics, Columbia University, New York, NY, USA

## Abstract

The habitable zone (HZ) is commonly defined as the range of distances from a host star within which liquid water, a key requirement for life, may exist on a planet's surface. Substantially more $CO_2$ than present in Earth's modern atmosphere is required to maintain clement temperatures for most of the HZ, with several bars required at the outer edge. However, most complex aerobic life on Earth is limited by $CO_2$ concentrations of just fractions of a bar. At the same time, most exoplanets in the traditional HZ reside in proximity to M dwarfs, which are more numerous than Sun-like G dwarfs but are predicted to promote greater abundances of gases that can be toxic in the atmospheres of orbiting planets, such as carbon monoxide (CO). Here we show that the HZ for complex aerobic life is likely limited relative to that for microbial life. We use a 1D radiative-convective climate and photochemical models to circumscribe a Habitable Zone for Complex Life (HZCL) based on known toxicity limits for a range of organisms as a proof of concept. We find that for $CO_2$ tolerances of 0.01, 0.1, and 1 bar, the HZCL is only 21%, 32%, and 50% as wide as the conventional HZ for a Sun-like star, and that CO concentrations may limit some complex life throughout the entire HZ of the coolest M dwarfs. These results cast new light on the likely distribution of complex life in the universe and have important ramifications for the search for exoplanet biosignatures and technosignatures.

## 1. Introduction

The search for habitable environments and life beyond our solar system is a deeply compelling scientific goal, as evidenced by a focus on these areas in the recent National Academies of Sciences report on Exoplanet Science Strategy (National Academies of Sciences, Engineering, and Medicine 2018). To date, over 3900 exoplanets have been discovered, some of which may possess conditions amendable for the emergence and maintenance of planetary biospheres (e.g., Anglada-Escudé et al. 2016; Ribas et al. 2016; Turbet et al. 2016; Gillon et al. 2017; Wolf 2017; Lincowski et al. 2018; Meadows et al. 2018). Discussions of the search for life beyond the solar system often begin with the circumstellar habitable zone (HZ)—the predicted range of distances from a star within which a planet with an $N_2$–$CO_2$–$H_2O$ atmosphere and a climate system stabilized by carbonate-silicate feedback can maintain surface temperatures conducive to the presence of liquid water (Walker et al. 1981; Kasting et al. 1993; Kopparapu et al. 2013). As conventionally defined, the inner edge of the HZ (IHZ) is delimited by the incident stellar flux above which a runaway (or moist) greenhouse occurs, while the outer edge of the HZ (OHZ) is determined by the "maximum greenhouse," an upper limit on the extent to which additional atmospheric $CO_2$ can compensate for decreasing stellar flux (Figure 1).

Crucially, the HZ is regarded as the future starting point for spectroscopic biosignature searches because temperate surfaces allow for significant exchange of gases between the (potential) biosphere and atmosphere (Kasting et al. 2014; Schwieterman et al. 2018). As a result, the occurrence rates of exoplanets in the HZ are of high scientific interest (Kopparapu 2013; Petigura et al. 2013; Dressing & Charbonneau 2015; Kane et al. 2016). Moreover, potential biosignature yields for future flagship space telescopes are based on our current understanding of the HZ (Stark et al. 2014, 2015; Kopparapu et al. 2018).

The basic requirement for surface liquid water is predicated on a subset of the minimum conditions needed for a simple, microbial biosphere, and limited attention has been paid to the conditions that may limit more complex life. We focus here primarily on potential limitations to higher metazoan (animal) analogs, encompassing relatively large (millimeter- to meter-scale), tissue-grade aerobic heterotrophs with blood vascular (circulatory) systems, although many of our results are applicable to diploblastic organisms, like sponges, that rely entirely on diffusion of $O_2$. However, we note that the term "complex life" is often applied to a wider variety of organisms on Earth, including plants and fungi, any of which may also be limited by the chemical consequences predicted by the underlying HZ concept. It is also important to acknowledge that the evolutionary template afforded by life on Earth is not necessarily inclusive of all of the evolutionary pathways that could exist on other worlds, which may provide for unknown compensatory mechanisms for living with extreme conditions prohibitive for complex life as we know it on Earth. Nevertheless, we find it a useful starting point to consider the possible distribution of familiar forms of biological complexity —in a manner analogous to the practical assumption of water-based life required by the conventional HZ.

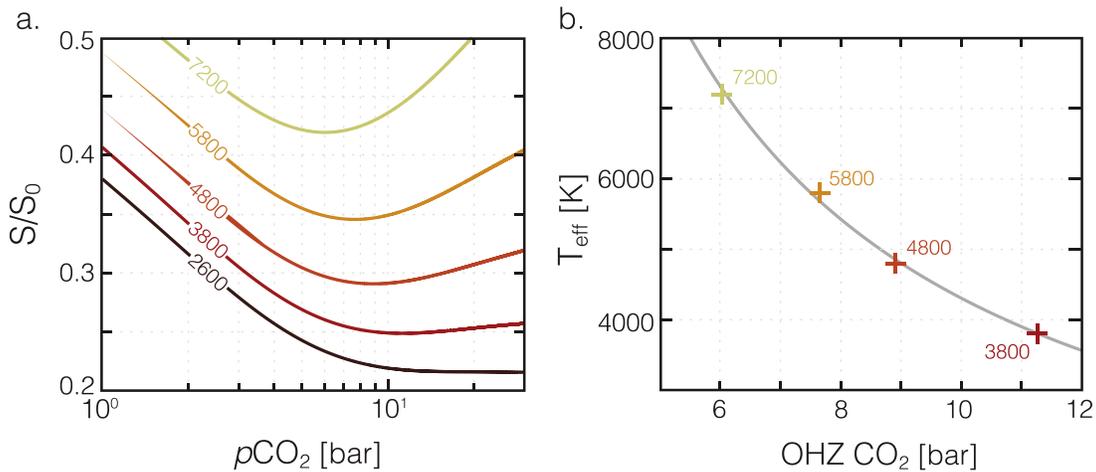

**Figure 1.** Estimated atmospheric $CO_2$ at the outer edge of the Habitable Zone (OHZ). Shown in (a) are fits to a series of 1D radiative-convective climate models (Kopparapu et al. 2013) in which the effective stellar flux ($S/S_0$) required to maintain a surface temperature of 273 K is computed as a function of atmospheric $pCO_2$ for a range of stellar hosts (with stellar effective temperatures denoted for each curve). The minimum $S/S_0$ value for each curve corresponds to the atmospheric $pCO_2$ at the OHZ for each star. Shown in (b) are atmospheric $pCO_2$ values at the OHZ as a function of stellar effective temperature. The gray curve shows the power fit used to derive the ranges for atmospheric $pCO_2$ at the OHZ shown [$pCO_2 = aT_{eff}^b$; $a = 3.1563 \times 10^{-4}$; $b = -0.963$; $r^2 = 0.997$].

Some previous studies have analyzed the potential impact of temperature on complex life within the HZ (e.g., Silva et al. 2017), while Bounama et al. (2007) was one of the very few previous studies to consider possible $CO_2$ limits on plant and animal life in a coupled stellar and geochemical evolution model to estimate the prevalence of complex life in the galaxy. However, we are aware of no existing study that has reported formal numerical HZ boundaries for complex life based on possible limitations due to $CO_2$ or that has proposed CO or other toxic gases as potential biochemical constraints.

Here we explore potential limitations to complex life in the HZ as a starting point for more detailed explorations in the future. In Section 2, we compare the predicted $CO_2$ concentrations at the outer edge of the HZ for FGKM stars to known physiological limits for complex aerobic life on Earth. In Section 3, we use a 1D photochemical model to predict likely CO concentrations for Earth-twins orbiting FGKM stars and compare these to known limits for CO toxicity. In Section 4, we use a 1D radiative-convective climate model to predict limiting HZ boundaries for (progressively less conservative) $CO_2$ limitations of 0.01, 0.1, and 1 bar and combine this with our results from Section 3 to propose a "Habitable Zone for Complex Life" (HZCL). We conclude with some implications of our results in Section 5.

## 2. Requirements of Complex Life and $CO_2$ Levels in the HZ

The origin and diversification of complex life on Earth is fundamentally tied to the rise of oxygen ($O_2$) in our atmosphere and oceans (Lyons et al. 2014; Planavsky et al. 2014; Reinhard et al. 2016). Physical and geochemical evidence for eukaryotes (complex cells with organelles, such as mitochondria and chloroplasts) is absent from the rock record until well after the earliest accumulation of free $O_2$ in Earth's atmosphere ~2.3 billion years ago (Ga) (Knoll 2014; Luo et al. 2016), while the first metazoan (animal) life emerged only in the last ~700 million years (Erwin et al. 2011; Zumberge et al. 2018). Later, significant increases in biological complexity on Earth, such as the Cambrian Explosion (~542 Ma; Lee et al. 2013), occurred against the backdrop of a more strongly oxygenated planetary atmosphere.

The metabolic oxidation of organic matter with $O_2$ produces significantly more free energy than any other plausible respiratory process, and $O_2$ is the only high-potential oxidant sufficiently stable to accumulate within planetary atmospheres (Catling et al. 2005). As a result, it is likely that the centrality of molecular $O_2$ in the emergence and expansion of a complex biosphere on Earth is a general phenomenon (Catling et al. 2005). However, complex aerobic life can be strongly impacted by $CO_2$ and CO—the latter of which is produced by $CO_2$ photolysis and surface biological activity. Both are expected to be present in various concentrations throughout the HZ. Importantly, as of yet we have little predictive capacity for determining which inhabited planets may build up $O_2$ in their atmospheres as a function of basic observables like insolation. In contrast, the HZ paradigm makes specific predictions about the level of $CO_2$ (or other greenhouse gases) required to keep a surface temperate. Therefore, while $O_2$ is likely the primary driver for the origin, development, and survival of complex life, assumptions embedded in basic climate physics of the HZ concept and consequent chemistry provide a more immediately discernable way of circumscribing potential boundaries for complexity.

One of the fundamental assumptions underlying the conventional HZ is that the carbonate-silicate cycle, in which atmospheric $CO_2$ levels are regulated by the effect of temperature on $CO_2$ consumption during rock weathering, will act to modulate atmospheric $CO_2$ concentrations (and thus surface temperatures) as a function of insolation (Walker et al. 1981). Near the inner edge of the HZ, clement surface temperatures can be maintained at low $CO_2$ concentrations similar to those of the modern Earth (tens to hundreds of ppm), but for the middle and outer regions of the HZ, atmospheric $CO_2$ concentrations need to be much higher in order to maintain temperatures conducive for surface liquid water—up

**Table 1**
Sample Physiological $CO_2$ Limits and Phanerozoic Atmospheric $CO_2$ Ranges

| Description | Limit (ppmv)* | Limit (bar) |
|---|---|---|
| Human limits | | |
| OSHA PEL[a] | 5000 | 0.0051 |
| NIOSH/ACGIH PEL[b] | 5000 | 0.0051 |
| NIOSH/ACGIH STEL[b] | 30,000 | 0.0304 |
| CDC IDLH[b] | 40,000 | 0.0405 |
| Squid limits | | |
| "Likely" long-term lethal (1), (2) | 6700 | 0.0068 |
| Acutely lethal (1), (2) | 26,500 | 0.0269 |
| Teleost fish limits | | |
| Lower acute lethality (3), (4) (range) | 30,000 | 0.0304 |
| Upper acute lethality (3), (4) (range) | 50,000 | 0.0507 |
| Phanerozoic Limits | | |
| Observed low [ice core record] (5) | 190 | $1.93 \times 10^{-4}$ |
| Observed high [proxy record] (6) | 4000 | 0.0041 |
| Modeled high [GEOCARB] (7), (8) | 10,000 | 0.0101 |

**Notes.**
[a] https://www.osha.gov/dsg/annotated-pels/tablez-1.html
[b] https://www.cdc.gov/niosh/idlh/124389.html

**References.** (1) Reipschläger & Pörtner (1996); (2) Pörtner et al. (2004); (3) Hayashi et al. (2004); (4) Ishimatsu et al. (2008); (5) Galbraith & Eggleston (2017); (6) Royer (2014); (7) Royer et al. (2014); (8) Lenton et al. (2018).

**Table 2**
Assumed Range of Stellar Effective Temperature ($T_{eff}$) for the Spectral Classes Considered here, along with the Estimated Range of Atmospheric $CO_2$ near the Outer Edge of the Habitable Zone

| Spectral Class | Bound | Stellar $T_{eff}$ [K] | $CO_2$ Minimum bar |
|---|---|---|---|
| … | … | | |
| G | low | 5300 | 8.91 |
| | high | 6000 | 7.92 |
| K | low | 3900 | 11.94 |
| | high | 5200 | 9.07 |
| F | low | 6000 | 7.92 |
| | high | 7600 | 6.32 |
| M | low | 2300 | 19.74 |
| | high | 3800 | 12.24 |

to many bars approaching the outer edge (Figure 1). For example, coupled orbital and GCM studies of the ostensible HZ planet Kepler 62-f have found that 3–5 bars of $CO_2$ would be required to maintain clement surface conditions (Shields et al. 2016b), a value that is up to ~1000 times greater than any witnessed during the entire history of complex life on Earth (see Table 1).

Elevated $CO_2$ levels can impose severe physiological stress on complex aerobic organisms (Pörtner et al. 2004; Wittmann & Pörtner 2013). Physiological responses to elevated $CO_2$ (hypercapnia) can be complex—often interacting across molecular, cellular, and organismal scales (Azzam et al. 2010)—but are most often regulated by respiratory acidosis and associated changes to ion buffering in internal fluids (Permentier et al. 2017). High atmospheric $CO_2$ also alters oceanic chemistry by lowering marine pH, with deleterious impacts on calcifying organisms and organisms that cannot effectively buffer internal pH (Wittmann & Pörtner 2013; Goodwin et al. 2014; Bennett et al. 2017). Indeed, physiological stress at elevated $CO_2$ has been proposed as a significant causal factor in major mass extinctions on Earth, particularly in the ocean (Knoll et al. 2007; Clarkson et al. 2015). Many of these negative impacts are predicted to occur as a result of anthropogenic $CO_2$ emissions over the next century, the effects of which are dwarfed by predicted OHZ $CO_2$ abundances (Table 1; Figure 1).

We estimate the $CO_2$ required at the outer edge of the conventional HZ as follows. First, we calculate the atmospheric $CO_2$ values corresponding to the minimum $S_{eff}$ at stellar effective temperatures of 2600, 3800, 4800, 5800, and 7200 K by fitting polynomial expressions to results from an ensemble of 1D radiative-convective climate models as presented in Kopparapu et al. (2013). These values represent the conventional "maximum greenhouse limit" above which Rayleigh scattering by $CO_2$ will lead to decreasing surface temperatures even as atmospheric $CO_2$ increases. We then fit a power function to these discrete values for atmospheric $CO_2$ at minimum $S_{eff}$ in order to derive a continuous function for the location of the maximum greenhouse limit as a function of stellar effective temperature (Figure 1). We remove the 2600 K result because it is effectively asymptotic at high atmospheric $CO_2$, and because small errors in the polynomial fit at this effective temperature can lead to spurious results. Finally, we assume a range of effective temperature for each spectral class (Table 2) and use this range to estimate atmospheric $CO_2$ corresponding to the maximum greenhouse limit for each star.

We can translate these $CO_2$ levels into estimates of surface marine pH, which can also be limiting for complex organisms (Wittmann & Pörtner 2013; Goodwin et al. 2014; Bennett et al. 2016). Marine pH values are estimated based on atmospheric $pCO_2$ using the csys3 package (Zeebe & Wolf-Gladrow 2001). We use the total pH scale with equilibrium constants from Roy et al. (1993), with pressure corrections according to Millero (1995). All calculations assume $T = 25°C$ and $S = 35‰$ as well as a background pressure of 1 bar to which a given $CO_2$ pressure is added. We assume that on average the global ocean maintains saturation with respect to calcite ($\Omega_{cal} = 1$):

$$\Omega_{cal} = \frac{[Ca^{2+}][CO_3^{2-}]}{K_{sp}^{cal}}, \quad (1)$$

where brackets denote concentration and $K_{sp}^{cal}$ denotes the solubility product of calcite ($CaCO_3$) at ambient $T$, $S$, and $P$. Assuming saturation and a globally averaged marine $Ca^{2+}$ abundance in Equation (1) allows us to calculate $[CO_3^{2-}]$, which together with an assumed $pCO_2$ allows us to solve the full seawater carbonate system. We also assume a globally averaged total marine boron concentration ($[B(OH)_3] + [B(OH)_4^-]$) of 416 $\mu$mol kg$^{-1}$ (Zeebe & Wolf-Gladrow 2001). We note that assuming saturation with respect to aragonite (another polymorph of $CaCO_3$) yields a higher estimated pH at a given atmospheric $pCO_2$, but on the scale of our analysis this difference is minor.

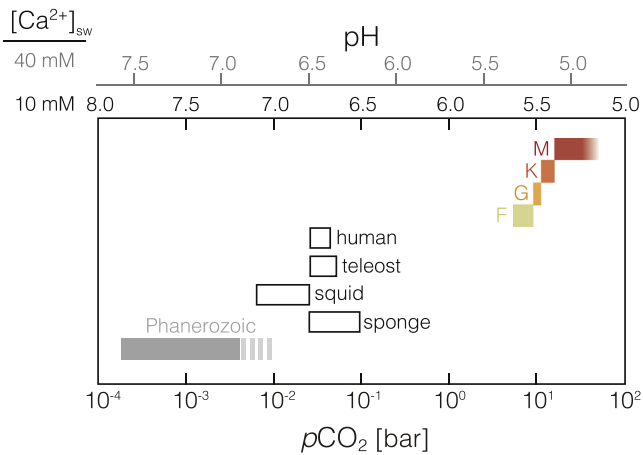

**Figure 2.** Physiological $CO_2$ limits at the outer edge of the habitable zone. Open bars show upper long-term physiological $CO_2$ tolerances from a range of complex organisms on Earth (Table 1). Gray bar shows the range of atmospheric $pCO_2$ levels during the Phanerozoic (540 million years ago to the present) according to geochemical proxies (solid)) and time-dependent biogeochemical models (dashed). Colored symbols show estimated values for $pCO_2$ at the outer edge of the habitable zone for F-, G-, K-, and M-type (red) stars. Upper scale shows marine pH assuming dissolved $Ca^{2+}$ concentrations of 10 and 40 mmol kg$^{-1}$. Note that because we assume saturation with respect to $CaCO_3$ in the surface ocean (rather than slight oversaturation as is characteristic of Earth's modern surface ocean) our pH value for roughly modern atmospheric $pCO_2$ is slightly below that observed.

When we compare the predicted atmospheric $CO_2$ abundances for planets at the outer edge of the HZ for FGKM main-sequence stars (or equivalent pH values) to levels of acute lethality in a range of complex organisms (Figure 2), we find that predicted atmospheric $CO_2$ is three to four orders of magnitude greater than the highest values estimated for the last 500 million years on Earth. Furthermore, the predicted $CO_2$ at the OHZ boundary is one to two orders of magnitude greater than the upper limits for the most $CO_2$-tolerant complex organisms known. Commonly proposed alternative greenhouse gases for extending the HZ (Pierrehumbert & Gaidos 2011; Seager 2013; Ramirez & Kaltenegger 2017, 2018)—$CH_4$ and/or $H_2$—are strongly reducing and thus chemically incompatible at high concentrations with the levels of $O_2$ required for the energy-intensive metabolisms of large, complex organisms (Catling et al. 2005). As a result, the biological requirements for both high $O_2$ and low $CO_2$ suggest that the potential for the development and radiation of complex life is strongly curtailed within the extended HZ and is limited to only a portion of the traditional HZ.

### 3. CO Toxicity and Enhanced Photochemical Lifetimes for Late-type Stars

An additional obstacle to complex life may be found in high-$O_2$ atmospheres on planets orbiting late-type stars, where certain photochemical conditions can lead to relatively high atmospheric CO levels (Schwieterman et al. 2019). CO is a highly toxic gas for humans and other vertebrates because their oxygen-carrying biomolecule hemoglobin has orders of magnitude higher bonding affinity for CO than for $O_2$ (Ryter & Otterbein 2004). For humans, CO concentrations exceeding $\sim$100 ppm are lethal for exposure times of 8 hr or more; CO levels of several hundred to thousands of ppm are lethal in only tens of minutes (National Research Council 2010). Longer-term exposure limits for CO are substantially lower than this, and epidemiological studies have shown that transient CO levels as low as $\sim$1 ppm in urban areas are associated with poor health outcomes (Liu et al. 2018).

A planet with a high-$O_2$ atmosphere may accumulate harmful levels of CO as a result of direct or indirect production by the biosphere through photolysis of dissolved organic matter in the surface ocean (Fichot & Miller 2010), production by phytoplankton (Conte et al. 2019), or biomass burning (Andreae & Merlet 2001)—even if $CO_2$ levels are relatively low. For planets orbiting cool stars, a deficit of near-UV radiation results in substantially less OH production and greatly increased atmospheric lifetimes of CO, along with other important biogenic gases relevant for biosignature detection such as $CH_4$ (Segura et al. 2005; Harman et al. 2015, 2018; Rugheimer et al. 2015; Nava-Sedeño et al. 2016; Schwieterman et al. 2019).

To illustrate this problem, we use a 1D photochemical model to predict atmospheric CO abundances for Earth twins (78% $N_2$, 21% $O_2$, and 360 ppm $CO_2$) orbiting FGKM stars and compare them to human toxicity limits. We use the photochemical component of the publicly available Atmos[10] model. Atmos is derived from the photochemical code originally developed by the Kasting group (Kasting et al. 1979; Pavlov et al. 2001) but with several additions and modifications. Recently, the upgraded code has been used to calculate photochemically self-consistent atmospheres for the Archean Earth and hazy planets orbiting other stars (Arney et al. 2016, 2017) and for calculating self-consistent trace gas abundances in the atmosphere of Proxima Centauri b (Meadows et al. 2018).

We begin with a converged atmosphere for the modern Earth (1 bar, 78% $N_2$, 21% $O_2$ v/v, and 360 ppm $CO_2$) and use the model to calculate the CO concentration resulting from a net flux of $3 \times 10^{11}$ cm$^{-2}$ s$^{-1}$ (equivalent to 1280 Tg yr$^{-1}$), which is the CO flux required by the model to reproduce the empirical CO mixing ratio in Earth's modern atmosphere ($1.1 \times 10^{-7}$ v/v). This value compares favorably with empirically estimated terrestrial CO fluxes (745.67–1112.80 Tg yr$^{-1}$; Zhong et al. 2017), mostly generated by terrestrial biomass burning. The temperature–pressure and water vapor mixing ratio profiles are consistent with a planet with a surface temperature of 288 K.

We then alter the input stellar spectrum while maintaining this surface flux to calculate the resulting CO abundance, a procedure employed by similar studies (Segura et al. 2005; Rugheimer et al. 2013; Meadows et al. 2018). The stellar spectra used by the code are identical to those used in prior studies (Segura et al. 2005, 2003; Arney et al. 2017; Lincowski et al. 2018; Meadows et al. 2018) and consist of empirical reconstructions of measured stellar spectra, including atomic and molecular absorption features. These stellar spectra are available online in the VPL stellar spectral database[11] and are built into the Atmos codebase. Our list of photochemical boundary conditions is given in Table 3. We also calculated the resulting CO concentrations when multiplying this CO flux by factors of 0.1, 0.33, 3, and 10 to circumscribe a range of plausible CO concentrations on a world with a terrestrial biosphere and an oxygen-rich atmosphere at the inner edge of a

---

[10] https://github.com/VirtualPlanetaryLaboratory/atmos
[11] https://depts.washington.edu/naivpl/content/spectral-databases-and-tools

**Table 3**
Boundary Conditions for Modeling Atmospheric CO Abundance for HZ Planets around a Range of Stellar Hosts in *Atmos*

| Chemical Species[1] | Deposition Velocity (cm s$^{-1}$) | Flux (molecules cm$^{-2}$ s$^{-1}$) | Mixing Ratio |
|---|---|---|---|
| O | 1 | ... | ... |
| $O_2$ | ... | ... | 0.21 |
| $N_2$ | ... | ... | 0.78 |
| $CO_2$ | ... | ... | $3.6 \times 10^{-4}$ |
| $H_2O$ | ... | ... | fixed[2] |
| H | 1 | ... | ... |
| OH | 1 | ... | ... |
| $HO_2$ | 1 | ... | ... |
| $H_2$ | $2.4 \times 10^{-4}$ | ... | $5.3 \times 10^{-7}$ |
| CO | $(0-1.2) \times 10^{-4}$ | variable | ... |
| HCO | 1 | ... | ... |
| $H_2CO$ | 0.2 | ... | ... |
| $CH_4$ | 0 | variable | ... |
| $CH_3$ | 1 | ... | ... |
| NO | $3 \times 10^{-4}$ | $1 \times 10^9$ | ... |
| $NO_2$ | $3 \times 10^{-3}$ | ... | ... |
| HNO | 1 | ... | ... |
| $H_2S$ | ... | $1 \times 10^8$ | ... |
| $SO_2$ | 1 | $1 \times 10^9$ | ... |
| $H_2SO_4$ | 1 | ... | ... |
| HSO | ... | ... | ... |
| $O_3$ | 0.07 | ... | ... |
| $HNO_3$ | 0.2 | ... | ... |
| $N_2O$ | ... | ... | $3.1 \times 10^{-7}$ |
| $HO_2NO_2$ | 0.2 | ... | ... |

**Notes.**
[1] Species included in the photochemical scheme with a deposition velocity and flux of 0 include: $C_2H_6$, HS, S, SO, $S_2$, $S_4$, $S_8$, $SO_3$, OCS, $S_3$, N, $NO_3$, and $N_2O_5$.
[2] The $H_2O$ profile is fixed to an Earth average.

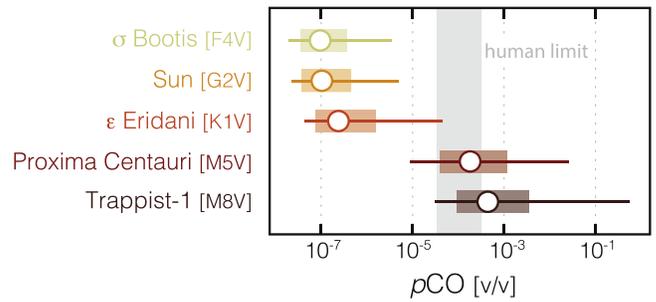

**Figure 3.** Steady-state atmospheric mixing ratios of CO for Earth-like planets around a range of stellar hosts. Open circles show results for the modern net surface CO flux ($3 \times 10^{11}$ molecules cm$^{-2}$ s$^{-1}$), while ranges show results of increasing/decreasing this flux by a factor of 3 (shaded bars) or 10 (horizontal lines). Also shown is the range between the short-term (1 hr; 330 ppm) and long-term (~40 ppm) permissible exposure limits for humans (NIOSH 2005). Calculations are performed assuming an atmospheric $pCO_2$ of 360 ppm (~$3.6 \times 10^{-4}$ bar), consistent with $CO_2$ levels predicted for the IHZ (both OHZ $CO_2$ and CO levels will be higher). Note the log scale.

star's HZ (Figure 3; Table 4). Some of these results were previously presented in Schwieterman et al. (2019) in the context of predicting the spectroscopic detectability of CO on inhabited exoplanets.

Our results show that CO concentrations on planets in the traditional HZ of FGK stars are unlikely to reach known toxicity limits for humans, at least in oxygen-rich atmospheres. However, for stars with effective temperatures below about 3200 K—for example, Proxima Centauri and TRAPPIST-1—we predict that CO concentrations can reach and significantly exceed short- and long-term human exposure limits. Because CO lifetimes are driven by destruction timescales set by OH (in $O_2$-rich atmospheres), dry planets orbiting interior to the traditional HZ may be problematic for complex life even around FGK stars (Abe et al. 2011; Zsom et al. 2013). In addition, cooler worlds with lower tropospheric $H_2O$ contents may have even higher CO concentrations than those shown here (Grenfell et al. 2007; Gao et al. 2015).

## 4. A Habitable Zone for Complex Life (HZCL)

Combining the potential impacts of high atmospheric $CO_2$ and the potential for abundant CO around cooler host stars, we use a 1D radiative-convective climate model (also a component of the *Atmos* package) to estimate the position of an illustrative "HZCL." For each case, we assume a 1-bar bulk atmosphere composed of 78% $N_2$, 21% $O_2$, and 1% Ar, with additional $CO_2$ pressures (in bar) of 0.01, 0.1, and 1. We consider these

$CO_2$ partial pressures to encompass conservative and optimistic ranges for long-term $CO_2$ limitation of complex aerobic life at 0.2 bar $O_2$ (i.e., Figure 2)—barring any currently unknown physiological mechanism for mitigating long-term hypercapnia at extremely high $CO_2$.

Our model calculations also assume a saturated troposphere, Earth's modern $O_3$ profile, and a 200 K stratosphere. (Note that the HZ boundaries are dependent on the choice of stratospheric temperatures (Ramirez 2018b); we use the temperature of the modern Earth because this is most consistent with stratospheric heating from the ozone.) The surface albedo is set to $A = 0.316$, which is tuned to reproduce the modern Earth's average surface temperature with a modern atmospheric composition (Kopparapu et al. 2013).

Assuming the above parameters, we use the climate model to calculate the effective stellar flux ($S_{eff}$) reaching the top of the planet's atmosphere required to warm the planetary surface to 273 K. We further derive a relationship between $S_{eff}$ and stellar effective temperature, following Kopparapu et al. (2013):

$$S_{eff} = S_{eff\odot} + aT_* + bT_*^2 + cT_*^3 + dT_*^4, \quad (2)$$

where $T_* = T - 5780$ K and coefficients for each scenario are listed in Table 5. These $S_{eff}$ values can be converted into distances by using the following equation:

$$d = \left(\frac{L/L_\odot}{S_{eff}}\right)^{0.5}, \quad (3)$$

where $L/L_\odot$ is the bolometric luminosity of the host star normalized by the Sun's bolometric luminosity. Figure 4 compares our analytic fits from Equation (2) to the conservative traditional HZ boundaries, which define the inner habitable zone (IHZ) by the "moist greenhouse" limit and the outer habitable zone (OHZ) by the "maximum greenhouse" limit.

We find that the HZCL is significantly restricted relative to the conventional HZ, even assuming an extremely high physiological $CO_2$ limit of 1 bar. Figure 5 illustrates the combined impact of physiological limitations of $CO_2$ and CO with our climate and photochemical results. We estimate that physiological $CO_2$ thresholds of 0.01, 0.1, and 1 bar correspond to HZCLs that are only 21%, 32%, and 50% as wide as the conventional HZ for a Sun-like star, with slightly smaller HZs for stars with lower effective temperatures. Our photochemical

Table 4
Estimated Atmospheric CO Concentrations around Select Stars as a Function of Surface CO Flux ($F_{CO}$), Scaled to that of the Modern Earth (e.g., $F_{CO} = 1$ for the Modern Earth Flux of $3 \times 10^{11}$ cm$^{-2}$ s$^{-1}$)

| Star | Spec Type | 0.1 x $F_{CO}$ | 0.33 x $F_{CO}$ | 1 x $F_{CO}$ | 3 x $F_{CO}$ | 10 x $F_{CO}$ |
|---|---|---|---|---|---|---|
| Sigma Boötis | F4V | $2.15 \times 10^{-8}$ | $3.75 \times 10^{-8}$ | $9.60 \times 10^{-8}$ | $3.91 \times 10^{-7}$ | $3.67 \times 10^{-6}$ |
| Sun | G2V | $2.36 \times 10^{-8}$ | $4.14 \times 10^{-8}$ | $1.10 \times 10^{-7}$ | $4.64 \times 10^{-7}$ | $5.32 \times 10^{-6}$ |
| Epsilon Eridani | K1V | $4.87 \times 10^{-8}$ | $8.71 \times 10^{-8}$ | $2.57 \times 10^{-7}$ | $1.65 \times 10^{-6}$ | $4.60 \times 10^{-5}$ |
| Prox. Centauri | M5V | $1.02 \times 10^{-5}$ | $4.28 \times 10^{-5}$ | $2.03 \times 10^{-4}$ | $1.25 \times 10^{-3}$ | $2.63 \times 10^{-2}$ |
| TRAPPIST-1 | M8V | $3.46E \times 10^{-5}$ | $1.13 \times 10^{-4}$ | $4.93 \times 10^{-4}$ | $4.04 \times 10^{-3}$ | 0.618 |

**Note.** CO Concentrations below 1 ppm are shaded in green, values between 1 and 100 ppm are shaded in yellow, and values above 100 ppm are shaded in red.

Table 5
Coefficients for Polynomial Expressions Bounding the Habitable Zone for Complex Life (HZCL) as Shown in Figure 4

| | 0.01 bar $CO_2$ | 0.1 bar $CO_2$ | 1 bar $CO_2$ | Maximum Greenhouse* |
|---|---|---|---|---|
| $S_{\mathrm{eff}\odot}$ | 0.7658 | 0.6743 | 0.5478 | 0.3438 |
| a | $2.9282 \times 10^{-5}$ | $2.9876 \times 10^{-5}$ | $3.7339 \times 10^{-5}$ | $5.8942 \times 10^{-5}$ |
| b | $-4.9812 \times 10^{-9}$ | $-4.6788 \times 10^{-9}$ | $-4.6185 \times 10^{-9}$ | $1.6558 \times 10^{-9}$ |
| c | $-1.7743 \times 10^{-12}$ | $-1.9425 \times 10^{-12}$ | $-2.3907 \times 10^{-12}$ | $-3.0045 \times 10^{-12}$ |
| d | $-3.2974 \times 10^{-17}$ | $-1.0263 \times 10^{-16}$ | $-8.7570 \times 10^{-17}$ | $-5.2983 \times 10^{-16}$ |

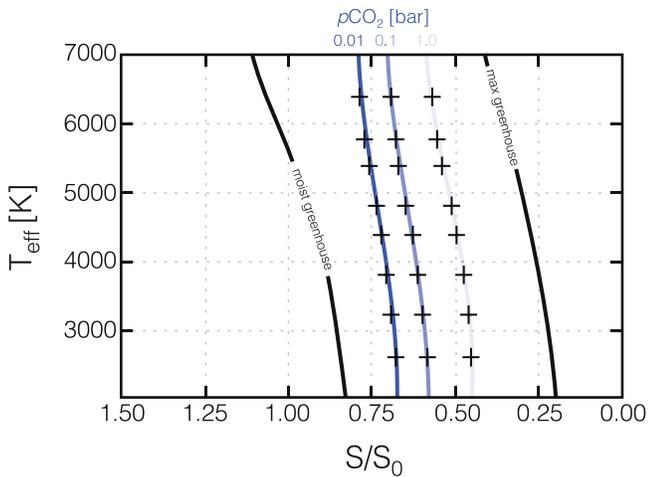

**Figure 4.** Polynomial fits to individual runs of our 1D radiative-convective climate model. Crosses show individual model runs, while green curves show polynomial fits using the coefficients given in Table 5. Also shown for reference are the moist greenhouse and maximum greenhouse limits on the conventional habitable zone from Kopparapu et al. (2013). The polynomial fits to individual model runs are shown in Figure 5.

model results suggest that high CO concentrations may limit some forms of complex life for host stars with $T_{\mathrm{eff}} < 3200$ K. These results further suggest that the regions of habitability available for complex life (as it know it on Earth) are significantly smaller than conventional circumstellar HZs.

## 5. Discussion

Given variability in biochemical response across metazoan clades and frequent secondary adaptation to high $CO_2$ conditions, it is challenging to prescribe a specific limiting $CO_2$ value for complex life in the HZ. Nevertheless, a recent review focusing on anthropogenically elevated $CO_2$ impacts on ocean ecosystems found universally negative impacts on extant marine life, including corals, echinoderms, mollusks, crustaceans, and fishes at $pCO_2 \geqslant 1\%$ (Wittmann & Pörtner 2013). Notably, most marine species experience negative repercussions from anthropogenically increasing $pCO_2$ at values much lower than this, within hundreds of ppm of the current concentration of $CO_2$ in our atmosphere (Wittmann & Pörtner 2013). Certain fishes have higher $CO_2$ tolerances, but most or all experience acute lethality at $pCO_2 < 5\%–10\%$ (Pörtner et al. 2004; Baker et al. 2015). Certain specialist animals, particularly burrowing mammals, can survive at higher $pCO_2$ for short periods (Shams et al. 2005; Kim et al. 2011). However, these responses represent secondary (rather than basal) adaptation. Specifically, survival of vertebrate animals at high $pCO_2$ ($\geqslant 10\%$) is a result of exceptional internal buffering capacities that must have evolved over time and were not present in the simplest and earliest animals, such as sponges. Crucially, the conditions necessary for the origin of complexity and its survival over geologic time are the most relevant for predicting the limits of the HZCL. In any case, even the highest long-term $CO_2$ exposure limits for secondarily adapted animals on Earth ($pCO_2$ ~10%) considerably restrict the HZCL relative to the conventional HZ (Figure 5), suggesting that $pCO_2$ limitations should be considered when estimating the distribution of complex life in the universe.

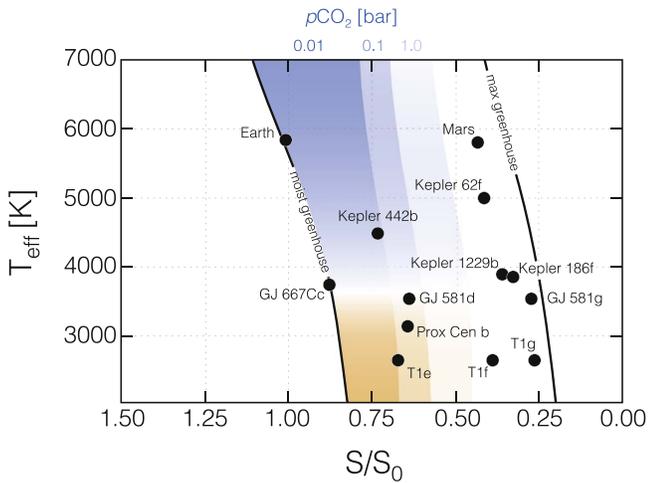

**Figure 5.** Stellar flux ($S/S_\odot$) boundaries for the traditional conservative HZ (Kopparapu et al. 2013) and the "Habitable Zone for Complex Life (HZCL)" assuming limiting $CO_2$ concentrations of 0.01 bar (dark blue), 0.1 bar (lighter blue), and 1 bar (lightest blue). The orange boundaries represent the low stellar effective temperatures where photochemical conditions may enhance CO lifetimes above the short-term permissible limits for humans (>100 ppm) at the IHZ, assuming a net surface molecular flux of $3 \times 10^{11}$ molecules cm$^{-2}$ s$^{-1}$. The positions of several known exoplanets within the HZ have also been plotted.

It is also important to consider that main-sequence stars brighten with time (Bahcall et al. 2001; Claire et al. 2012), leading to significant temporal shifts in the position of the HZ (and HZCL; Rushby et al. 2013). For example, the Sun's luminosity was ~70% of current levels when it entered the main sequence 4.6 billion years ago, and we are now about half of the way through its hydrogen-burning lifetime (Gough 1981; Bahcall et al. 2001). Because of such stellar brightening, planets that form within the traditional HZ but outside the HZCL will eventually enter the HZCL as their host star evolves along the main sequence and baseline atmospheric $CO_2$ is gradually reduced. A consequence of this relationship is that complex forms of life inhabiting a planet orbiting a main-sequence star must originate in the context of declining (high to low) $CO_2$ with geologic time and therefore cannot "adapt" to high $CO_2$ levels on a planetary scale. This trajectory is most problematic if there is a step on the evolutionary ladder of complexity that requires relatively neutral ocean pH or otherwise low $pCO_2$. Furthermore, Earth's current position in the HZCL at a distance amenable for complex and intelligent life is not tuned, though we live in a special time after the carbonate-silicate cycle has driven $CO_2$ to low values, but before the brightening Sun produces uninhabitable temperature conditions (Wolf & Toon 2015) or puts Earth into a moist or runaway greenhouse (Kasting 1988).

Lastly, we have shown here and elsewhere that CO concentrations are likely to be high on Earth-like planets orbiting late-type stars with similar surface molecular fluxes of CO (Schwieterman et al. 2019). This result is consistent with the seminal findings of Segura et al. (2005), who showed that the low NUV radiation from stars with low effective temperatures would act to drastically reduce the generation of OH radicals in Earth-like atmospheres on planets orbiting these stars, with the consequence that trace gases like $CH_4$ otherwise destroyed by OH could build up to high abundances. This prediction has been replicated in a multitude of other studies focused on assessing possible biosignature abundances and detectability on terrestrial planets orbiting late-type stars (e.g., Grenfell et al. 2013; Rugheimer et al. 2015; Lincowski et al. 2018; Meadows et al. 2018; Arney 2019; Wunderlich et al. 2019). An inescapable corollary to the prediction of high concentrations of biosignature gases, however, is that the concentrations of abiotic and biologically produced toxic trace gases like CO must also be high if their photochemical lifetimes are primarily set by interactions with OH radicals or more generally by NUV photons penetrating into the troposphere. Other toxic gases that may reach high concentrations on planets orbiting M dwarfs include volcanic emissions (e.g., $H_2S$), products of $NO_x$ chemistry (e.g., $NO_2$), and perhaps others. In addition to possible limits from toxic gas buildup, there are other habitability concerns with M dwarfs that may also apply to microbial life, such as potential atmospheric erosion from flares and climatic impacts from tidal locking (see, e.g., Shields et al. 2016a for a review), all of which may render M dwarf planets poor candidates for the development of complex or intelligent life.

## 6. Conclusions

Our results have a number of important implications for the search for exoplanet biosignatures and complex life beyond our solar system. For example, our predictions of a more limited zone for complex life place constraints on the planetary environments suitable for the evolution of intelligence, if it requires free $O_2$ and limited concentrations of $CO_2$, CO, and other potentially toxic trace gases. One implication is that we may not expect to find remotely detectable signs of intelligent life ("technosignatures") on planets orbiting late M dwarfs or on potentially habitable planets near the outer edge of their HZs. These $CO_2$ and CO limits should be considered in future targeted SETI searches (Tarter 2001, 2004; Turnbull & Tarter 2003). The possible importance of photochemistry in creating environments conducive to complex and intelligent life further suggests a strong need for stellar UV characterization (e.g., France et al. 2016; Loyd et al. 2016; Youngblood et al. 2016) not only for biosignature prediction and assessment but also for SETI target prioritization.

Furthermore, the need for significant greenhouse warming from reduced gases should rule out complex aerobic life, as well as remotely detectable $O_2$ as a biosignature, from a large region of the expanded HZ (Seager 2013; Ramirez 2018a). More broadly, limitations on complex life by $CO_2$ and CO may partially address why we find ourselves near the inner edge of the HZ of a G-dwarf star rather than near the center or toward the outer edge of the HZ around one of the much more numerous M-dwarf stars (Waltham 2017; Haqq-Misra et al. 2018), as this condition is most favorable from the perspectives of both $CO_2$ drawdown and limited toxic gas abundance.

Moving forward, it will be critical to use coupled 3D climate-photochemical models to more accurately circumscribe the HZCL (e.g., Chen et al. 2018). Estimates of the conventional HZ boundaries have been shown to differ between 1D and 3D models due to a range of factors, including the impact of spatially variable surface albedo, atmospheric mass, surface gravity, rotation rate, continental area and distribution, and orbital parameters (Kopparapu et al. 2016; Wolf 2017). Trace greenhouse gases like $N_2O$ and $CH_4$, in addition to pressure broadening from inert gases such as $N_2$, will also impact these boundaries (e.g., Vladilo et al. 2013; Kopparapu et al. 2014). Nevertheless, our results highlight the importance of a planet's relative HZ location and atmospheric photochemistry in constraining the planetary potential for complex life. We suggest that the expected physiological impacts of high $CO_2$, CO, and other gases possibly toxic for complex life should be considered in attempts to search for biological complexity beyond our solar system.

This work was supported by the NASA Astrobiology Institute Alternative Earths team under Cooperative Agreement Number NNA15BB03A and the VPL under Cooperative Agreement Number NNA13AA93A. The VPL is also supported by the NASA Astrobiology Program under grant number 80NSSC18K0829. E.W.S. is additionally grateful for support from the NASA Postdoctoral Program, administered by the Universities Space Research Association. S.L.O. acknowledges support from the T.C. Chamberlin postdoctoral fellowship in the Department of Geophysical Sciences at the University of Chicago. C.E.H gratefully acknowledges research support for the ROCKE-3D team through NASA's Nexus for Exoplanet System Science (NExSS), via solicitation NNH13ZDA017C. We thank both an anonymous referee and Roger Buick for helpful and critical comments that allowed us to improve our paper.

## ORCID iDs

Edward W. Schwieterman  https://orcid.org/0000-0002-2949-2163

Stephanie L. Olson  https://orcid.org/0000-0002-3249-6739

Chester E. Harman  https://orcid.org/0000-0003-2281-1990